\pdfoutput=1
\documentclass[12pt,a4paper]{article}

\usepackage{ifthen} % for conditional statements
\newboolean{pdflatex}
\setboolean{pdflatex}{true} % False for eps figures 

\newboolean{articletitles}
\setboolean{articletitles}{true} % False removes titles in references

\newboolean{uprightparticles}
\setboolean{uprightparticles}{false} %True for upright particle symbols

\newboolean{inbibliography}
\setboolean{inbibliography}{false} %True once you enter the bibliography

\usepackage{overpic}

\usepackage[top=1in, bottom=1.25in, left=1in, right=1in]{geometry}

\usepackage{microtype}
\usepackage{lineno}  % for line numbering during review
\usepackage{xspace} % To avoid problems with missing or double spaces after
\usepackage{caption} %these three command get the figure and table captions automatically small

\usepackage{graphicx}  % to include figures (can also use other packages)
\usepackage{color}
\usepackage{colortbl}
\graphicspath{{./figs/}} % Make Latex search fig subdir for figures

\usepackage{amsmath} % Adds a large collection of math symbols
\usepackage{amssymb}
\usepackage{amsfonts}
\usepackage{upgreek} % Adds in support for greek letters in roman typeset

\newcommand*\patchAmsMathEnvironmentForLineno[1]{%
\expandafter\let\csname old#1\expandafter\endcsname\csname #1\endcsname
\expandafter\let\csname oldend#1\expandafter\endcsname\csname
end#1\endcsname
 \renewenvironment{#1}%
   {\linenomath\csname old#1\endcsname}%
   {\csname oldend#1\endcsname\endlinenomath}%
}
\newcommand*\patchBothAmsMathEnvironmentsForLineno[1]{%
  \patchAmsMathEnvironmentForLineno{#1}%
  \patchAmsMathEnvironmentForLineno{#1*}%
}
\AtBeginDocument{%
\patchBothAmsMathEnvironmentsForLineno{equation}%
\patchBothAmsMathEnvironmentsForLineno{align}%
\patchBothAmsMathEnvironmentsForLineno{flalign}%
\patchBothAmsMathEnvironmentsForLineno{alignat}%
\patchBothAmsMathEnvironmentsForLineno{gather}%
\patchBothAmsMathEnvironmentsForLineno{multline}%
\patchBothAmsMathEnvironmentsForLineno{eqnarray}%
}

\usepackage{hyperref}    % Hyperlinks in references
\usepackage[all]{hypcap} % Internal hyperlinks to floats.

\def\lhcb {\mbox{LHCb}\xspace}

\def\MagUp {\mbox{\em Mag\kern -0.05em Up}\xspace}

\ifthenelse{\boolean{uprightparticles}}%
{

 \def\Pmu         {\ensuremath{\upmu}\xspace}

 \def\Ppi         {\ensuremath{\uppi}\xspace}

 \def\PDelta      {\ensuremath{\Delta}\xspace}                 
 \def\PXi      {\ensuremath{\Xi}\xspace}                 
 \def\PLambda      {\ensuremath{\Lambda}\xspace}                 
 \def\PSigma      {\ensuremath{\Sigma}\xspace}                 
 \def\POmega      {\ensuremath{\Omega}\xspace}                 
 \def\PUpsilon      {\ensuremath{\Upsilon}\xspace}                 
 
 \def\PB      {\ensuremath{\mathrm{B}}\xspace}                 
                  
 \def\PD      {\ensuremath{\mathrm{D}}\xspace}

 \def\PK      {\ensuremath{\mathrm{K}}\xspace}

 \def\Pb      {\ensuremath{\mathrm{b}}\xspace}

 \def\Pi      {\ensuremath{\mathrm{i}}\xspace}

 \def\Ps      {\ensuremath{\mathrm{s}}\xspace}

}
{

 \def\Pmu         {\ensuremath{\mu}\xspace}

 \def\Ppi         {\ensuremath{\pi}\xspace}

 \mathchardef\PDelta="7101
 \mathchardef\PXi="7104
 \mathchardef\PLambda="7103
 \mathchardef\PSigma="7106
 \mathchardef\POmega="710A
 \mathchardef\PUpsilon="7107
                  
 \def\PB      {\ensuremath{B}\xspace}                 
                  
 \def\PD      {\ensuremath{D}\xspace}

 \def\PK      {\ensuremath{K}\xspace}

 \def\Pb      {\ensuremath{b}\xspace}

 \def\Pi      {\ensuremath{i}\xspace}

 \def\Ps      {\ensuremath{s}\xspace}

}

\makeatletter
\ifcase \@ptsize \relax% 10pt
  \newcommand{\miniscule}{\@setfontsize\miniscule{4}{5}}% \tiny: 5/6
\or% 11pt
  \newcommand{\miniscule}{\@setfontsize\miniscule{5}{6}}% \tiny: 6/7
\or% 12pt
  \newcommand{\miniscule}{\@setfontsize\miniscule{5}{6}}% \tiny: 6/7
\fi
\makeatother

\DeclareRobustCommand{\optbar}[1]{\shortstack{{\miniscule (\rule[.5ex]{1.25em}{.18mm})}
  \\ [-.7ex] $#1$}}

\def\mup        {{\ensuremath{\Pmu^+}}\xspace}
\def\mun        {{\ensuremath{\Pmu^-}}\xspace} % muon negative (\mum is taken)
\def\mumu       {{\ensuremath{\Pmu^+\Pmu^-}}\xspace}

\def\squark    {{\ensuremath{\Ps}}\xspace}

\def\bquark    {{\ensuremath{\Pb}}\xspace}

\def\pion   {{\ensuremath{\Ppi}}\xspace}

\def\pim    {{\ensuremath{\pion^-}}\xspace}

\def\kaon    {{\ensuremath{\PK}}\xspace}
  \def\Kbar    {{\kern 0.2em\overline{\kern -0.2em \PK}{}}\xspace}

\def\KorKbar    {\kern 0.18em\optbar{\kern -0.18em K}{}\xspace}

\def\Kp      {{\ensuremath{\kaon^+}}\xspace}

  \def\Dbar    {{\kern 0.2em\overline{\kern -0.2em \PD}{}}\xspace}

\def\DorDbar    {\kern 0.18em\optbar{\kern -0.18em D}{}\xspace}

\def\B       {{\ensuremath{\PB}}\xspace}
\def\Bbar    {{\ensuremath{\kern 0.18em\overline{\kern -0.18em \PB}{}}}\xspace}

\def\BorBbar    {\kern 0.18em\optbar{\kern -0.18em B}{}\xspace}
\def\Bz      {{\ensuremath{\B^0}}\xspace}

\def\Bd      {{\ensuremath{\B^0}}\xspace}
\def\Bs      {{\ensuremath{\B^0_\squark}}\xspace}
\def\Bsb     {{\ensuremath{\Bbar{}^0_\squark}}\xspace}

  \def\Y#1S{\ensuremath{\PUpsilon{(#1S)}}\xspace}% no space before {...}!

\def\Lz          {{\ensuremath{\PLambda}}\xspace}
\def\Lbar        {{\ensuremath{\kern 0.1em\overline{\kern -0.1em\PLambda}}}\xspace}
\def\LorLbar    {\kern 0.18em\optbar{\kern -0.18em \PLambda}{}\xspace}

\def\Lb      {{\ensuremath{\Lz^0_\bquark}}\xspace}

\newcommand{\decay}[2]{\ensuremath{#1\!\to #2}\xspace}         % {\Pa}{\Pb \Pc}

\def\to                 {\ensuremath{\rightarrow}\xspace}

\def\BdToKpi      {\decay{\Bd}{\Kp\pim}}

\def\AT#1     {\ensuremath{A_{\mathrm{T}}^{#1}}\xspace}           % 2

\def\Bsmm     {\decay{\Bs}{\mup\mun}}
\def\Bdmm     {\decay{\Bd}{\mup\mun}}

\def\C#1      {\ensuremath{\mathcal{C}_{#1}}\xspace}                       % 9
\def\Cp#1     {\ensuremath{\mathcal{C}_{#1}^{'}}\xspace}                    % 7
\def\Ceff#1   {\ensuremath{\mathcal{C}_{#1}^{\mathrm{(eff)}}}\xspace}        % 9  
\def\Cpeff#1  {\ensuremath{\mathcal{C}_{#1}^{'\mathrm{(eff)}}}\xspace}       % 7
\def\Ope#1    {\ensuremath{\mathcal{O}_{#1}}\xspace}                       % 2
\def\Opep#1   {\ensuremath{\mathcal{O}_{#1}^{'}}\xspace}                    % 7

\newcommand{\tev}{\ifthenelse{\boolean{inbibliography}}{\ensuremath{~T\kern -0.05em eV}\xspace}{\ensuremath{\mathrm{\,Te\kern -0.1em V}}}\xspace}
\newcommand{\gev}{\ensuremath{\mathrm{\,Ge\kern -0.1em V}}\xspace}
\newcommand{\mev}{\ensuremath{\mathrm{\,Me\kern -0.1em V}}\xspace}
\newcommand{\kev}{\ensuremath{\mathrm{\,ke\kern -0.1em V}}\xspace}
\newcommand{\ev}{\ensuremath{\mathrm{\,e\kern -0.1em V}}\xspace}
\newcommand{\gevc}{\ensuremath{{\mathrm{\,Ge\kern -0.1em V\!/}c}}\xspace}
\newcommand{\mevc}{\ensuremath{{\mathrm{\,Me\kern -0.1em V\!/}c}}\xspace}
\newcommand{\gevcc}{\ensuremath{{\mathrm{\,Ge\kern -0.1em V\!/}c^2}}\xspace}
\newcommand{\gevgevcccc}{\ensuremath{{\mathrm{\,Ge\kern -0.1em V^2\!/}c^4}}\xspace}
\newcommand{\mevcc}{\ensuremath{{\mathrm{\,Me\kern -0.1em V\!/}c^2}}\xspace}

\def\invfb   {\ensuremath{\mbox{\,fb}^{-1}}\xspace}

\def\ps   {\ensuremath{{\rm \,ps}}\xspace}

\newcommand{\chisqip}{\ensuremath{\chi^2_{\rm IP}}\xspace}

\def\gsim{{~\raise.15em\hbox{$>$}\kern-.85em
          \lower.35em\hbox{$\sim$}~}\xspace}
\def\lsim{{~\raise.15em\hbox{$<$}\kern-.85em
          \lower.35em\hbox{$\sim$}~}\xspace}

\def\sPlot{\mbox{\em sPlot}\xspace}
\def\tell1  {TELL1\xspace}
\def\ukl1   {UKL1\xspace}

\usepackage{cite} % Allows for ranges in citations
\usepackage{mciteplus}

\newcommand{\BRof}[1]{\ensuremath{{\cal B}(#1)}\xspace}
\newcommand{\Bsmumu}{\ensuremath{B^0_s \to\mu^+\mu^-}\xspace}
\newcommand{\Bdmumu}{\ensuremath{B^0\to\mu^+\mu^-}\xspace}
\newcommand{\Bmm}{\ensuremath{B^{0}_{(s)}\to\mu^+\mu^-}\xspace}
\newcommand{\Bmumu}{\ensuremath{B^0_{(s)}\to \mu^+\mu^-}\xspace}

\def\bdkpi{\ensuremath{B^0 \to K^+ \pi^-}\xspace}

\newcommand{\bpimumu}{\ensuremath{B^{0(+)} \to \pi^{0(+)} \mu^+ \mu^-}\xspace}
\newcommand{\BdPiMuNu}{\ensuremath{\ensuremath{B^0}\to \pi^- \mu^+ \nu_\mu}\xspace}

\newcommand{\BhMuNu}{\ensuremath{\ensuremath{B^0_{(s)}}\to h^- \mu^+ \nu_\mu}\xspace}

\newcommand{\BuJpsiK}{\ensuremath{B^+\to J/{\psi}K^+}\xspace}
\def\bujpsik{\BuJpsiK}

\newcommand{\Bhh}{\ensuremath{B^0_{(s)}\to h^+{h}^{\prime -}}\xspace}

\newcommand{\bbdim}{\ensuremath{b\bar{b}\to \mu^+ \mu^- X}\xspace}
\newcommand{\BsJpsiPhi}{\ensuremath{B^0_s\to J/\psi \phi}\xspace}
\newcommand{\BdKpi}{\ensuremath{B^0\to K^+\pi^-}\xspace}
\newcommand{\BsKK}{\ensuremath{B^0_s\to K^+K^-}\xspace}
\def\Lbpmunu{\ensuremath{\Lb \to p \mu^- \bar \nu_{\mu}}\xspace} 
\newcommand{\BcJpsiMuNu}{\ensuremath{B^+_c\to J/\psi\mu^+\nu_\mu}\xspace}
\newcommand{\Bsd}{\ensuremath{B^0_{(s)}}\xspace}

 \def\Bsbrshort{\ensuremath{\left(3.0\pm 0.6^{\,+0.3}_{\,-0.2}\right)\times 10^{-9}}\xspace}

\def\ys{\ensuremath{y_s}\xspace}
\def\ADeltaGamma{\ensuremath{A^{\mu^+\mu^-}_{\Delta\Gamma}}\xspace}
\def\Bdobslimitnf{\ensuremath{3.4\times 10^{-10}}\xspace} %  95% CL 

\def\Bdsigma{\ensuremath{1.6\,\sigma}\xspace}

\def\Bssigmafirst{\ensuremath{7.8}\xspace}

\def\BorBbars    {\kern 0.18em\optbar{\kern -0.18em B}}

\def\Bdbr{\ensuremath{\left(1.5^{\,+1.2\,+0.2}_{\,-1.0\,-0.1}\right)\times 10^{-10}}\xspace}

\newcommand{\CLs}{\ensuremath{\textrm{CL}_{\textrm{s}}}\xspace}

\def\Y#1S{\ensuremath{\Upsilon{(#1S)}}\xspace}% no space before {...}!

\def\invfb      {\ensuremath{\mbox{\,fb}^{-1}}\xspace}

\def\BDT{BDT\xspace}
\newcommand{\comment}[1]{}

\definecolor{darkred}{rgb}{0.6,0.0,0.0}
\definecolor{darkgreen}{rgb}{0.0,0.5,0.0}
\definecolor{lightgreen}{rgb}{0.75,1.0,0.75}
\definecolor{lightred}{rgb}{1.0,0.75,0.75}
\definecolor{lightblue}{rgb}{0.75,0.75,1.0}
\definecolor{darkblue}{RGB}{100,100,200}
\definecolor{verylightblue}{rgb}{0.9,0.9,1.0}
\definecolor{verylightred}{rgb}{1.0,0.9,0.9}
\definecolor{lightgray}{rgb}{0.9,0.9,0.9}
\definecolor{verylightgray}{rgb}{0.95,0.95,0.95}
\definecolor{darkgray}{rgb}{0.75,0.75,0.75}
\definecolor{orange}{rgb}{1.0,0.75,0.0}

\usepackage{longtable} % only for template; not usually to be used in PAPERs
\usepackage{multirow}
\usepackage{afterpage}

\begin{document}

%%%%%%%%%%%%%%%%%%%%%%%%%
%%%%% Title     %%%%%%%%%
%%%%%%%%%%%%%%%%%%%%%%%%%
\renewcommand{\thefootnote}{\fnsymbol{footnote}}
\setcounter{footnote}{1}

% %%%%%%% CHOOSE TITLE PAGE--------
%\onecolumn
% \input{title-LHCb-ANA}
%\input{title-LHCb-CONF}
% ===============================================================================
% Purpose: LHCb-PAPER journal paper title page template
% Author: 
% Created on: 2010-09-25
% ===============================================================================

%%%%%%%%%%%%%%%%%%%%%%%%%
%%%%%  TITLE PAGE  %%%%%%
%%%%%%%%%%%%%%%%%%%%%%%%%
\begin{titlepage}
\pagenumbering{roman}

% Header ---------------------------------------------------
\vspace*{-1.5cm}
\centerline{\large EUROPEAN ORGANIZATION FOR NUCLEAR RESEARCH (CERN)}
\vspace*{1.5cm}
\hspace*{-0.5cm}
\begin{tabular*}{\linewidth}{lc@{\extracolsep{\fill}}r}
\ifthenelse{\boolean{pdflatex}}% Logo format choice
{\vspace*{-2.7cm}\mbox{\!\!\!\includegraphics[width=.14\textwidth]{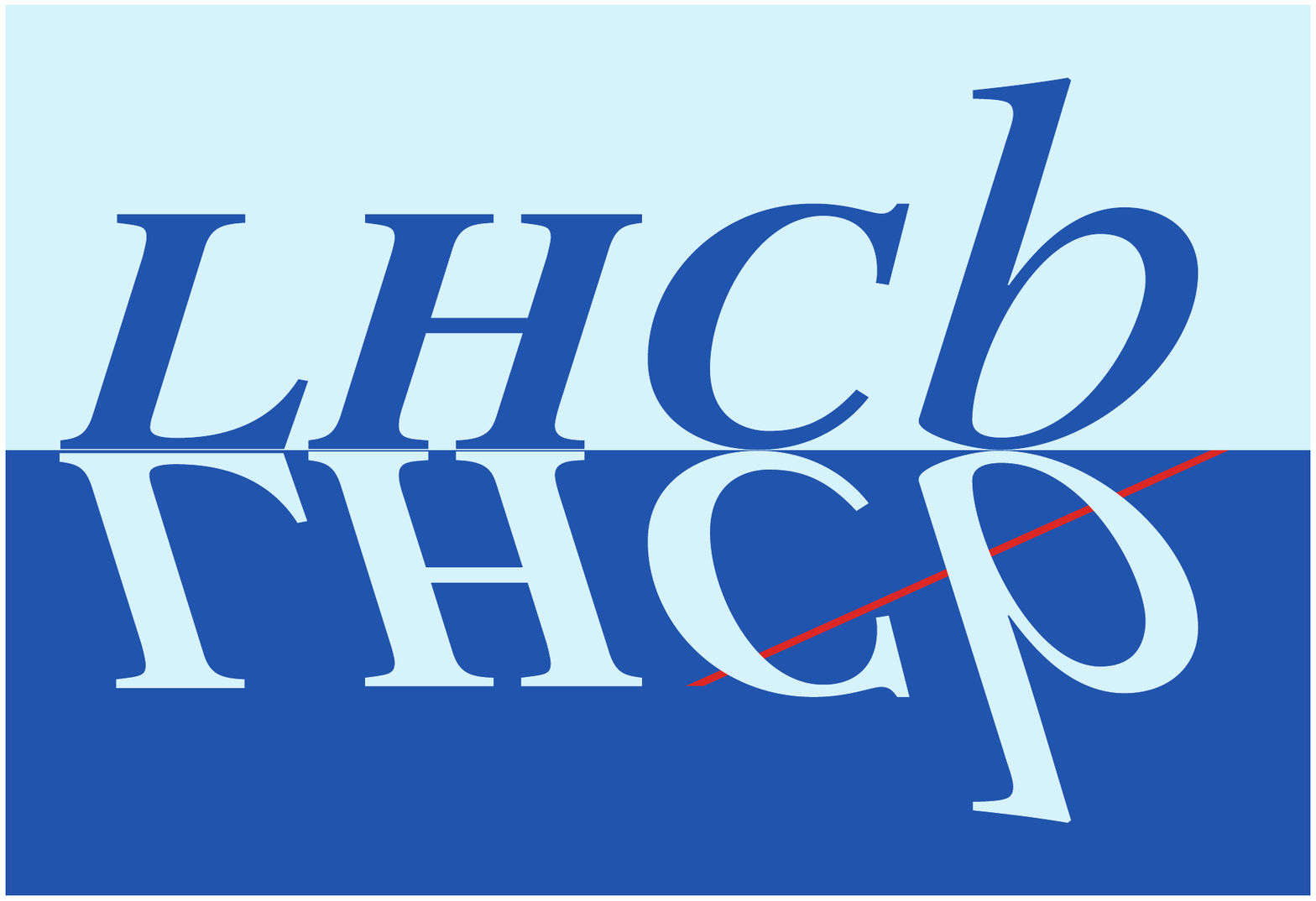}} & &}%
{\vspace*{-1.2cm}\mbox{\!\!\!\includegraphics[width=.12\textwidth]{lhcb-logo.eps}} & &}%
\\
 & & CERN-EP-2017-041 \\  % ID 
 & & LHCb-PAPER-2017-001 \\  % ID 
%& & \today \\ 
& & March 16, 2017 \\ 
 & & \\
\end{tabular*}

%\vspace*{3.0cm}
\vspace*{1.2cm}

{\bf\boldmath\huge
  \begin{center}
    Measurement of the $B^0_s\to\mu^+\mu^-$ branching fraction and effective lifetime and search for $B^0\to\mu^+\mu^-$ decays
\end{center}
}

\vspace*{0.8cm}

\begin{center}
The LHCb collaboration\footnote{Authors are listed at the end of this paper.}
\end{center}

\vspace{\fill}

\begin{abstract}
  \noindent
A search for the rare decays \Bsmumu and \Bdmumu is performed at the LHCb experiment using data collected in $pp$ collisions 
corresponding to a total integrated luminosity of 4.4\invfb. An excess of \Bsmumu decays is observed with a significance of \Bssigmafirst standard deviations, representing the first observation of this decay in a single experiment. 
The branching fraction is measured to be ${\cal B}(\Bsmumu)=\Bsbrshort$, where the first uncertainty is statistical and the second systematic. The first measurement of the \Bsmumu effective lifetime, \mbox{$\tau(\Bsmm)=2.04\pm 0.44\pm 0.05 \ps$}, is reported. 
No significant excess of \Bdmumu decays is found and a 95\% confidence level upper limit, \mbox{$\BRof \Bdmumu < \Bdobslimitnf$}, is determined. All results are in agreement with the Standard Model expectations.
\end{abstract}

\vspace*{1.0cm}

\begin{center}
Published in Phys. Rev. Lett. 118 (2017), 191801
\end{center}

\vspace{\fill}

{\footnotesize 
\centerline{\copyright~CERN on behalf of the \lhcb collaboration, licence \href{http://creativecommons.org/licenses/by/4.0/}{CC-BY-4.0}.}}
\vspace*{2mm}

\end{titlepage}

\newpage
\setcounter{page}{2}
\mbox{~}

\cleardoublepage

%\twocolumn
% %%%%%%%%%%%%% ---------

\renewcommand{\thefootnote}{\arabic{footnote}}
\setcounter{footnote}{0}

%%%%%%%%%%%%%%%%%%%%%%%%%%%%%%%%
%%%%%  Table of Content   %%%%%%
%%%%%%%%%%%%%%%%%%%%%%%%%%%%%%%%
%%%% Uncomment next 2 lines if desired
%\tableofcontents
%\cleardoublepage

%%%%%%%%%%%%%%%%%%%%%%%%%
%%%%% Main text %%%%%%%%%
%%%%%%%%%%%%%%%%%%%%%%%%%

\pagestyle{plain} % restore page numbers for the main text
\setcounter{page}{1}
\pagenumbering{arabic}
%\linenumbers

%% Uncomment during review phase. 
%% Comment before a final submission.
%\linenumbers

% You can include short sections directly in the main tex file.
% However, for larger papers it is desirable to split the text into
% several semiautonomous files, which can be revised independently.
% This is especially useful when developing a document in
% collaboration with several people, since then different parts can be
% edited independently.  This type of file organization is shown here.
% 

\sloppy

\noindent Within the Standard Model (SM) of particle physics, the \Bdmm and \Bsmm decays are very rare, because they only occur through loop diagrams and are helicity-suppressed. Since they are characterised by a purely leptonic final state, and thanks to the progress in lattice QCD calculations~\cite{Witzel:2013sla,Na:2012kp,Bazavov:2011aa}, their time-integrated branching fractions, \BRof \Bsmumu = \mbox{$(3.65 \pm 0.23) \times 10^{-9}$} and \BRof \Bdmumu = \mbox{$(1.06 \pm 0.09) \times 10^{-10}$}~\cite{Bobeth:2013uxa}, are predicted in the SM with small uncertainty. 
These features make the \Bmm decays sensitive probes for physics beyond the SM, for example an extended Higgs sector~\cite{Babu:1999hn,Isidori:2001fv,Buras:2002wq}. 
The measurement of these processes has attracted considerable theoretical and experimental interest, culminating in the recent observation of the \Bsmm decay and evidence of the \Bdmm decay reported by the LHCb and CMS collaborations~\cite{LHCb-PAPER-2014-049}. This has been obtained by combining their datasets collected in $pp$ collisions in 2011 and 2012~\cite{LHCb-PAPER-2013-046,Chatrchyan:2013bka}. The measured branching fractions, \BRof \Bsmumu = \mbox{$(2.8^{+0.7}_{-0.6})\times 10^{-9}$} and \BRof \Bdmumu = \mbox{$(3.9^{+1.6}_{-1.4})\times 10^{-10}$}, are consistent with SM predictions. The ATLAS collaboration has also recently reported a search for these decays~\cite{Aaboud:2016ire}.

In the $\Bs-\Bsb$ system, the light and heavy mass eigenstates are characterised by 
a sizable difference between their decay widths, $\Delta\Gamma=0.082\pm 0.007\ps^{-1}$~\cite{Amhis:2016xyh}. In the SM, only the heavy state decays to $\mu^+\mu^-$, but this condition does not necessarily hold in New Physics scenarios~\cite{DeBruyn:2012wk}. The contributions from the two states can be disentangled by measuring the \Bsmm effective lifetime, which, in the search for physics beyond the SM, is a complementary probe to the branching fraction measurement.
The effective lifetime is defined as $\tau_{\mu^+\mu^-}\equiv \int^\infty_0 t\,\Gamma\!\left(B_s(t)\to\mu^+\mu^- \right)dt/\int^\infty_0 \Gamma\!\left(B_s(t)\to\mu^+\mu^- \right) dt$, 
where $t$ is the decay time of the \Bs or \Bsb meson and $\Gamma\!\left(B_s(t)\to\mu^+\mu^- \right)\equiv\Gamma\!\left(\Bs(t)\to \mu^+\mu^-\right)+\Gamma\!\left(\Bsb(t)\to\mu^+\mu^-\right)$. The relation~\cite{DeBruyn:2012wj}
\begin{equation}\label{eq:tau_deltaagamma}
\tau_{\mu^+\mu^-}=\frac{\tau_{\Bs}}{1 - \ys^2} \left[ \frac{1+ 2 \ADeltaGamma \ys + \ys^2}{1 + \ADeltaGamma \ys} \right],
\end{equation}
holds, where $\tau_{\Bs}=1.510\pm0.005\ps$ is the \Bs mean lifetime and $y_s\equiv\tau_{\Bs}\Delta\Gamma/2=0.062\pm0.006$~\cite{Amhis:2016xyh,Olive:2016xmw}.
The parameter \ADeltaGamma is defined as $\ADeltaGamma=-2\Re(\lambda)/(1+|\lambda|^2)$, with \mbox{$\lambda=(q/p)(A(\Bsb\to\mu^+\mu^-)/A(\Bs\to\mu^+\mu^-))$}. The complex coefficients $p$ and $q$ define the mass eigenstates of the $\Bs-\Bsb$ system in terms of the flavour eigenstates~(see, e.g., Ref.~\cite{Amhis:2016xyh}), and $A(\Bs\to\mu^+\mu^-)$ ($A(\Bsb\to\mu^+\mu^-))$ is the $\Bs$ ($\Bsb$) decay amplitude. In the SM the quantity \ADeltaGamma is equal to unity but can assume any value in the range $[-1,1]$ in New Physics scenarios. 

This Letter reports measurements of the \Bsmm and \Bdmm time-integrated branching fractions, which supersede the previous LHCb results~\cite{LHCb-PAPER-2013-046}, and the first measurement of the \Bsmm effective lifetime. 
Results are based on data collected with the LHCb detector, corresponding to an integrated luminosity of 1\invfb  of $pp$ collisions at a centre-of-mass energy $\sqrt{s}=7\tev$, 2\invfb at $\sqrt{s}=8\tev$ and 1.4\invfb recorded at $\sqrt{s}=13\tev$. 
The first two datasets are referred to as Run 1 and the latter as Run 2.

At various stages of the analysis multivariate classifiers are employed to select the signal. In particular, after trigger and loose selection requirements, \Bmm candidates are classified according to their dimuon mass and the output variable, BDT, of a multivariate classifier based on a boosted decision tree~\cite{bdt_book}, which is employed to separate signal and combinatorial background. 
The signal yield is determined from a fit to the dimuon mass distribution of candidates and is converted into a branching fraction using as normalisation modes the decays \bdkpi and \bujpsik, with $J/{\psi}\to \mu^+ \mu^-$ (inclusion of charge-conjugated processes is implied throughout this Letter).

The analysis strategy is similar to that employed in Ref.~\cite{LHCb-PAPER-2013-046} and has been optimised to enhance the sensitivity to both \Bs\ and \Bd\ decays to \mumu. This is achieved through a better rejection of misidentified $b$-hadron decays such as \Bhh (where $h^{(\prime)} = \pi, K$) and the development of an improved boosted decision tree for the BDT classifier. The \Bsmumu effective lifetime is measured from the background-subtracted decay-time distribution of signal candidates in the lowest-background BDT region as defined later. To avoid potential biases, candidates in the dimuon mass signal region (\mbox{$[5200,5445]\mevcc$}) were not examined until the analysis procedure was finalised.

The \lhcb detector is a single-arm forward spectrometer covering the pseudorapidity range \mbox{$2<\eta<5$}, described in detail in Refs.~\cite{Alves:2008zz,LHCb-DP-2014-002}. 
It includes a high-precision tracking system consisting of a silicon-strip vertex detector, surrounding the $pp$ interaction region, a large-area silicon-strip detector located upstream of a dipole magnet with a bending power of about 4\,Tm, and three stations of silicon-strip detectors and straw drift tubes placed downstream of the magnet. Particle identification is provided by two ring-imaging Cherenkov detectors, an 
electromagnetic and a hadronic calorimeter, and a muon system composed of alternating layers of iron and multiwire proportional chambers.
The simulated events used in this analysis are produced using the software described in Refs.~\cite{Sjostrand:2006za,*Sjostrand:2007gs,*LHCb-PROC-2010-056,*Lange:2001uf,*Agostinelli:2002hh,*Allison:2006ve,*LHCb-PROC-2011-006,Golonka:2005pn}.

Candidate events for signal and normalisation are selected by a hardware trigger followed by a software trigger~\cite{Aaij:2012me}. The \Bmm candidates are predominantly selected by single-muon and dimuon triggers. 
The \BuJpsiK candidates are selected in a very similar way, the only difference being a different dimuon mass requirement in the software trigger. Candidate \Bhh decays are used as control and normalisation channels.

The \Bmm candidates are reconstructed by combining two oppositely charged particles with transverse momentum with respect to the beam, $p_{\rm T}$, satisfying $0.25<p_{\rm T}<40\gevc$, momentum $p< 500\gevc$, and high-quality muon identification~\cite{Archilli:2013npa}. Compared to the previous analysis, the muon identification requirements are tightened such that the misidentified \Bhh background is reduced by approximately 50\%, while the signal efficiency decreases by about 10\%. The muon candidates are required to form a secondary vertex with a vertex-fit $\chi^2$ per degree of freedom smaller than 9 and separated from any primary $pp$ interaction vertex (PV) by a flight distance significance greater than 15. Only muon candidate tracks with $\chisqip>25$ for any PV are selected, where \chisqip is defined as the difference between the vertex-fit $\chi^2$ of the PV formed with and without the particle in question.
In the selection, \Bsd candidates must have a decay time less than $9\,\tau_{\Bs}$, $\chisqip<25$ with respect to the PV for which the \chisqip is minimal (henceforth called the \Bsd PV), $p_{\rm T} > 0.5 \gevc$ and a dimuon mass in the range $[4900, 6000]\mevcc$. 
A \Bsd candidate is rejected if either of the two candidate muons combined with any other oppositely charged muon candidate in the event has a mass within 30\mevcc of the $J/\psi$ mass~\cite{Olive:2016xmw}. The normalisation channels are selected with almost identical requirements to those applied to the signal sample. The \Bhh selection is the same as that of \Bmumu, except that the muon identification criteria are replaced with hadron identification requirements. The \BuJpsiK decay is reconstructed by combining a muon pair, consistent with a $J/\psi$ from a detached vertex, and a kaon candidate with $\chisqip>25$ for all PVs in the event. These selection criteria are completed by a loose requirement on the response of a multivariate classifier, described in Ref.~\cite{LHCb-PAPER-2012-007} and unchanged since then, applied to candidates in both signal and normalisation channels. The classifier takes as input quantities related to the direction of the \Bsd candidate, its impact parameter with respect to the \Bsd PV, the separation between the final-state tracks, and their impact parameters with respect to any PV. After the trigger and selection requirements 78\,241 signal candidates are found, which form the dataset for the subsequent branching fraction measurement.

The separation between signal and combinatorial background is achieved by means of the BDT variable, where the boosted decision tree is optimised using simulated samples of \Bsmm events for signal and of \bbdim events for background. The classifier combines information from the following input variables: $\sqrt{\Delta\phi^2+\Delta\eta^2}$, where $\Delta\phi$ and $\Delta\eta$ are the azimuthal angle and pseudorapidity differences between the two muon candidates; the minimum $\chi^2_{\rm IP}$ of the two muons with respect to the \Bsd PV; the angle between the \Bsd candidate momentum and the vector joining the \Bsd decay vertex and \Bsd PV; the \Bsd candidate vertex-fit $\chi^2$ and impact parameter significance with respect to the \Bsd PV.
In addition, two isolation variables are included, to quantify the compatibility of the other tracks in the event with originating from the same hadron decay as the signal muon candidates. Most of the combinatorial background is composed of muons originating from semileptonic $b$-hadron decays, in which other charged particles may be produced and reconstructed. The isolation variables are constructed to recognise these particles and differ in the type of tracks being considered: the first considers tracks that have been reconstructed both before and after the magnet, while the second considers tracks reconstructed only in the vertex detector. The isolation variables are determined  based on the proximity of the two muon candidates to the tracks of the event and are optimised using simulated \Bsmm and \bbdim events. The proximity of each muon candidate to a track is measured using a multivariate classifier that takes as input quantities such as the angular and spatial separation between the muon candidate and the track, the signed distance between the muon-track vertex and the \Bsd candidate or primary vertex, and the kinematic and impact parameter information of the track. 

The \BDT variable is constructed to be distributed uniformly in the range [0,1] for signal, and to peak strongly at zero for background. Its correlation with the dimuon mass is below 5\%. Compared to the multivariate classifier used in the previous measurement~\cite{LHCb-PAPER-2013-046}, the combinatorial background with ${\rm\BDT}>0.25$ is reduced by approximately 50\%, mainly due to the improved performance of the isolation variables.

The expected \Bmm \BDT distributions are determined from those of \mbox{\BdKpi} decays in data after correcting them for distortions  due to trigger and muon identification. An additional correction is made for the \Bs signal, assuming the SM prediction, to account for the difference between the \Bz and \Bsmumu lifetimes, which affects the BDT distribution. 
The mass distribution of the signal decays is described by a Crystal Ball function~\cite{Skwarnicki:1986xj}. The peak values for the \Bs and \Bd mesons are obtained from the mass distributions of \BsKK and \BdKpi samples, respectively. The mass resolutions as function of \mumu mass are determined with a power-law interpolation between the measured resolutions of charmonium and bottomonium resonances decaying into two muons. 
The Crystal Ball radiative tail is obtained from simulated \Bsmumu events~\cite{Golonka:2005pn}, which are smeared such that they reproduce the 23\mevcc mass resolution measured in data.

The signal branching fractions are measured with
\begin{equation*}
\BRof \Bmumu=\frac{{\cal B}_{\rm norm}  \,{\rm \epsilon_{\rm norm}}\,f_{\rm norm} }{ N_{\rm norm}\,{\rm \epsilon_{sig}} \,f_{d(s)} } \times N_{\Bmumu}\equiv\alpha^{\rm norm}_{\Bmumu} \times N_{\Bmumu},
\end{equation*}
where $N_{\Bmumu}$ is the number of observed signal decays, $N_{\rm norm}$ is the number of normalisation-channel decays (\BuJpsiK and \BdKpi), ${\cal B}_{\rm norm}$ is the corresponding branching fraction~\cite{Olive:2016xmw}, and ${\rm \epsilon_{sig}}$ (${\rm \epsilon_{\rm norm}}$) is the total efficiency for the signal (normalisation) channel.
The fraction $f_{d(s)}$ indicates the probability for a $b$ quark to fragment into a $B^0_{(s)}$ meson. Assuming $f_d=f_u$, the fragmentation probability $f_{\rm norm}$ for the $B^0$ and $B^+$ normalisation channel is set to $f_d$. The value of $f_s/f_d$ in $pp$ collision data at $\sqrt{s}=7\tev$ has been measured by LHCb to be $0.259 \pm 0.015$~\cite{fsfd}. 
The stability of $f_s/f_d$ at $\sqrt{s}=8\tev$ and $13\tev$ is evaluated by comparing the observed variation of the ratio of the efficiency-corrected yields of \BsJpsiPhi and \BuJpsiK decays. The effect of increased collision energy is found to be negligible for data at $\sqrt{s}=8\tev$ while a scaling factor of $1.068\pm 0.046$ is applied for data at $\sqrt{s}=13\tev$.

The efficiency ${\rm \epsilon_{sig(norm)}}$ includes the detector acceptance, trigger, reconstruction and selection efficiencies of the final-state particles. The acceptance, reconstruction and selection efficiencies are computed with samples of simulated events whose decay-time distributions are generated according to the SM prediction. The tracking and particle identification efficiencies are determined using control channels in data~\cite{LHCb-DP-2013-002,Anderlini:2202412}. 
The trigger efficiencies are evaluated with data-driven techniques~\cite{Tolk:1701134}.

The numbers of \BuJpsiK and \bdkpi decays are \mbox{$(1964.2\pm 1.5)\times 10^3$} and \mbox{$(31.3\pm 0.4)\times 10^3$}, respectively. The normalisation factors derived from the two channels are consistent. Taking correlations into account, their weighted averages are $\alpha^{\rm norm}_{\Bsmumu}=(5.7\pm 0.4)\times 10^{-11}$ and $\alpha^{\rm norm}_{\Bdmumu}=(1.60\pm 0.04)\times 10^{-11}$.
In the SM scenario, the analysed data sample is expected to contain an average of $62\pm 6$ \Bsmm and $6.7\pm 0.6$ \Bdmm decays in the full BDT range.

The combinatorial background is distributed almost uniformly over the mass range. In addition, the signal region and the low-mass sideband (\mbox{$[4900,5200]\mevcc$}) are populated by backgrounds from exclusive $b$-hadron decays, which can be classified in two categories. The first includes \Bhh, \BdPiMuNu, $B^0_s \to K^- \mu^+ \nu_{\mu}$, and \Lbpmunu decays, where one or two hadrons are misidentified as a muon. The \Bhh, $B^0 \to \pi^- \mu^+ \nu_{\mu}$ and \Lbpmunu branching fractions are taken from Refs.~\cite{Olive:2016xmw,LHCb-PAPER-2015-013}, while a theoretical estimate for $B^0_s \to K^- \mu^+ \nu_{\mu}$ is obtained from Refs.~\cite{Bouchard:2014ypa,Flynn:2015mha}. The mass and BDT distributions of these decays are determined from simulated samples after calibrating the $K\to\mu$, $\pi\to\mu$ and $p\to\mu$ momentum-dependent misidentification probabilities using control channels in data. An independent estimate of the \Bhh, \BdPiMuNu and $B^0_s \to K^- \mu^+ \nu_{\mu}$ background yields is obtained by fitting the mass spectrum of $\pi^+\mu^-$ or $K^+\mu^-$ combinations selected in data, and rescaling the yields according to the $\pi\to\mu$ or $K\to\mu$ misidentification probability. The difference with respect to the results from the first method is assigned as a systematic uncertainty.
The second category includes the decays \mbox{$B^+_c \to J/\psi \mu^+ \nu_{\mu}$}, with $J/\psi  \to \mu^+ \mu^-$, and \mbox{\bpimumu}, which have at least two muons in the final state. The rate of \mbox{$B^+_c \to J/\psi \mu^+ \nu_{\mu}$} decays is evaluated from Refs.~\cite{LHCb-PAPER-2014-050,LHCb-PAPER-2014-025}, 
while those of \mbox{\bpimumu} decays are obtained from Refs.~\cite{LHCb-PAPER-2015-035,Wang:2012ab}. 
The expected yields of all exclusive backgrounds are estimated using the decay \BuJpsiK as the normalisation channel, with the exception of the \Bhh decays, which are normalised to the mode \BdKpi. 
The contributions from \mbox{$B^0_s\to\mu^+\mu^-\gamma$} and \mbox{$B^0_s \to \mu^+ \mu^- \nu_{\mu} \bar \nu_{\mu}$} decays~\cite{Melikhov:2004mk,Bobeth:2013uxa,Aditya:2012im} have a negligible impact on the signal yield determination. The expected background yields with ${\rm \BDT}>0.5$ in the signal region are $2.9\pm 0.3$ \Bhh, $1.2\pm 0.2$ \BcJpsiMuNu, $0.7\pm 0.2$ \Lbpmunu and $0.80\pm 0.06$ \BhMuNu decays. 
The \bpimumu background is negligible. Except for the misidentified \Bhh decays, which populate the \Bd signal region, the other modes are mostly concentrated in the low-mass sideband.

The Run~1 and Run 2 datasets are each divided into five subsets based on bins in the BDT variable with boundaries $0.0$, $0.25$, $0.4$, $0.5$, $0.6$ and $1.0$. 
The \Bsmm and \Bdmm branching fractions are determined with a simultaneous unbinned maximum likelihood fit to the dimuon  mass distribution in each BDT bin of the two datasets. 
 The \Bsmm and \Bdmm fractional yields in each BDT bin and the parameters of the Crystal Ball functions that describe the shapes of the mass distributions are Gaussian-constrained according to their expected values and uncertainties. 
The combinatorial background in each BDT bin is parameterised with an exponential function, with a common slope parameter for all bins of a given dataset, while the yield is allowed to vary independently. The exclusive backgrounds are included as separate components in the fit. Their overall yields as well as the fractions in each BDT bin are Gaussian-constrained according to their expected values. Their mass shapes are determined from a simulation for each BDT bin.

\begin{figure}[t!]
\begin{center}
\includegraphics*[width=.9\columnwidth]{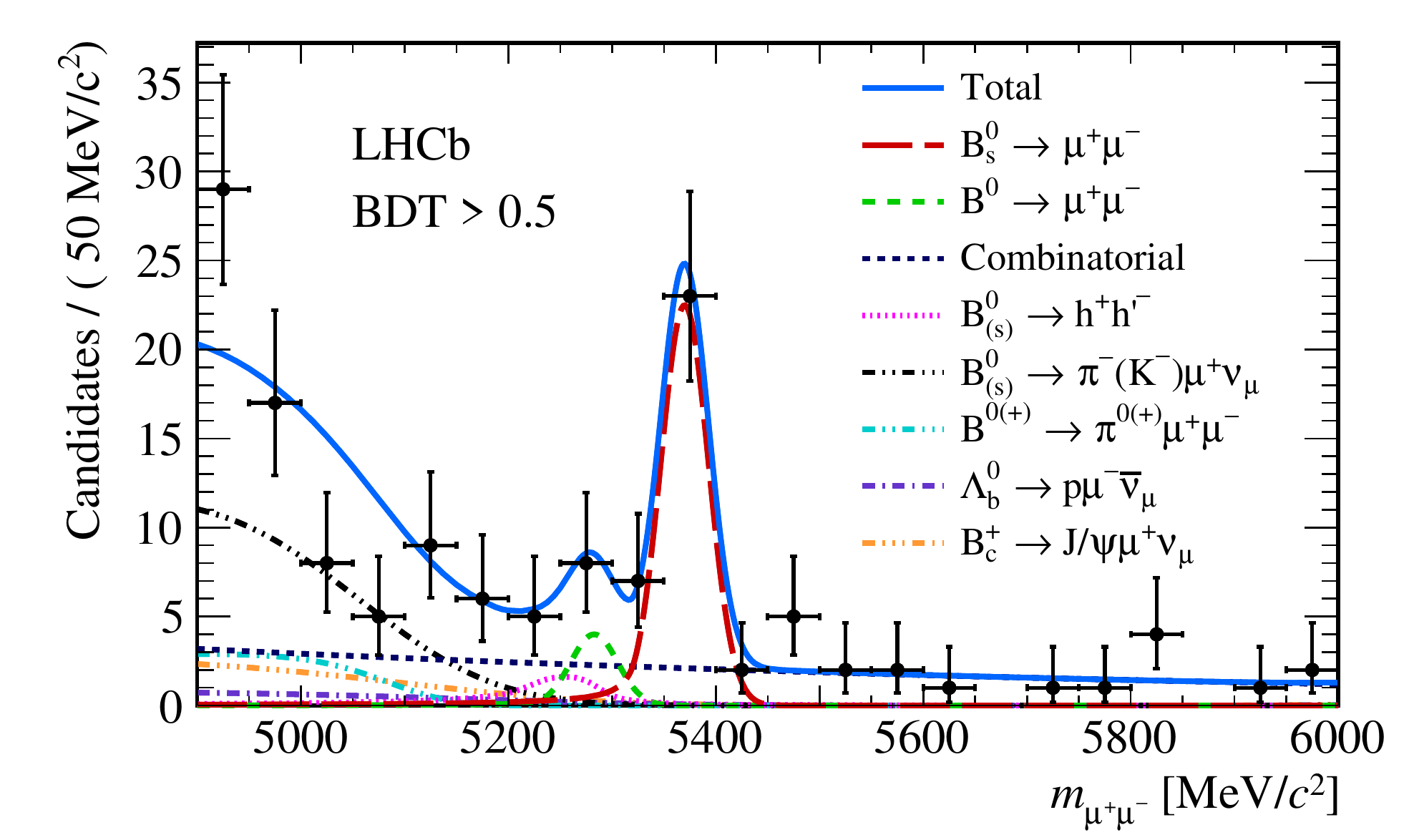}
\end{center}
\caption{\small Mass distribution of the selected \Bmm candidates (black dots) with ${\rm \BDT}>0.5$. The result of the fit is overlaid, and the different components are detailed.
}\label{fig:mass}
\end{figure}

The values of the \Bsmm and \Bdmm branching fractions obtained from the fit are \mbox{$\BRof \Bsmumu=\Bsbrshort$} and \mbox{$\BRof \Bdmumu=\Bdbr$}. 
The statistical uncertainty is derived by repeating the fit after fixing all the fit parameters, except the \Bdmumu and \Bsmumu branching fractions, the background yields and the slope of the combinatorial background, to their expected values. The systematic uncertainties of $\BRof \Bsmumu$  and $\BRof\Bdmumu$ are dominated by the uncertainty on $f_s/f_d$ and the knowledge of the exclusive backgrounds, respectively.
The correlation between the two branching fractions is negligible. 
The  mass distribution of the \mbox{\Bmm} candidates with ${\rm \BDT}>0.5$ is shown in Fig.~\ref{fig:mass}, together with the fit result~\cite{roofit}.

An excess of \Bsmumu candidates with respect to the expectation from background is observed with a significance of \Bssigmafirst standard deviations~($\sigma$), while the significance of the \Bdmumu signal is \Bdsigma. The significances are determined, using Wilks' theorem~\cite{Wilks:1938dza}, from the difference in likelihood between fits with and without the signal component.

Since no significant \Bdmm signal is observed, an upper limit on the branching fraction is set using the \CLs method~\cite{Read:2002hq}. The ratio between the likelihoods in two hypotheses, signal plus background and background only, is used as the test statistic. The likelihoods are computed with nuisance parameters fixed to their nominal values. Pseudo-experiments are used for the evaluation of the test statistic in which the nuisance parameters are floated according to their uncertainties. The resulting upper limit on \BRof \Bdmumu is \mbox{$\Bdobslimitnf$} at 95\% confidence level.

The selection efficiency and BDT distribution of \Bsmm decays depend on the lifetime, which in turn depends on the model assumption entering Eq.~\ref{eq:tau_deltaagamma}. This introduces a further model-dependence in the measured time-integrated branching fraction. In the fit, the SM value  $\tau(\Bsmumu)=\tau_{B_s}/(1-y_s)$ is assumed, corresponding to $\ADeltaGamma=1$. The model dependence is evaluated by repeating the fit under the $\ADeltaGamma=0$ and $-1$ hypotheses, finding an increase of the branching fraction with respect to the SM assumption of 4.6\% and 10.9\%, respectively. The dependence is approximately linear in the physically allowed \ADeltaGamma range.

For the \Bsmm lifetime determination, the data are background-subtracted with the \sPlot technique~\cite{Pivk:2004ty}, using a fit to the dimuon mass distribution to disentangle signal and background components statistically. Subsequently, a fit to the signal decay-time distribution is made with an exponential function multiplied by the acceptance function of the detector.
The \Bs candidates are selected using criteria similar to those applied in the branching fraction analysis, the main differences being a reduced dimuon mass window, $[5320,6000]\mevcc$, and looser particle identification requirements on the muon candidates. 
The former change allows the fit model for the \Bsmm signal to be simplified by removing most of the \Bdmm and exclusive background decays that populate the lower dimuon mass region, while the latter increases the signal selection efficiency. Furthermore, instead of performing a fit in bins of \BDT, a requirement of \BDT$>0.55$ is imposed. All these changes minimise the statistical uncertainty on the measured effective lifetime.
This selection results in a final sample of 42 candidates.

The mass fit includes the \Bsmm and combinatorial background components. 
The parameterisations of the mass shapes are the same as used in the branching fraction analysis. The correlation between the mass and the reconstructed decay time of the selected candidates is less than 3\%.
  
The variation of the trigger and selection efficiency with decay time is corrected for in the fit by introducing an acceptance function, determined from simulated signal events that are weighted to match the properties of the events seen in data. The use of simulated events to determine the decay-time acceptance function is validated by measuring the effective lifetime of \BdKpi decays selected in data. The measured effective lifetime is $1.52\pm 0.03 \ps$, where the uncertainty is statistical only, consistent with the world average~\cite{Olive:2016xmw}. The statistical uncertainty on the measured \BdKpi lifetime is taken as the systematic uncertainty associated with the use of simulated events to determine the \Bsmm acceptance function. 

The accuracy of the fit for the \Bsmm effective lifetime is estimated using a large number of simulated experiments with signal and background contributions equal, on average, to those observed in the data. The contamination from \Bdmm, \Bhh and semileptonic decays above 5320\mevcc is small and not included in the fit. The effect on the effective lifetime from the unequal production rate of \Bs and \Bsb mesons~\cite{LHCb-PAPER-2014-042} is negligible.
A bias may also arise if $\ADeltaGamma\neq\pm 1$, with the consequence that the underlying decay-time distribution is the sum of two exponential distributions with the lifetimes of the light and heavy mass eigenstates. In this case, as the selection efficiency varies with the decay time, the returned value of the lifetime from the fit is not exactly equal to the definition of the effective lifetime even if the decay-time acceptance function is correctly accounted for. This effect has been evaluated for the scenario where there are equal contributions from both eigenstates to the decay. 
The result can also be biased if the background has a much longer mean lifetime than \Bsmm decays; this is mitigated by an upper decay-time cut of 13.5\ps. Any remaining bias is evaluated using the background decay-time distribution of the much larger \BdToKpi data sample. 
All of these effects are found to be small compared to the statistical uncertainty and combine to give 0.05\ps, with the main contributions arising from the fit accuracy and the decay-time acceptance (0.03\ps each).
\begin{figure}[t]
\begin{center}
\includegraphics*[width=.9\columnwidth]{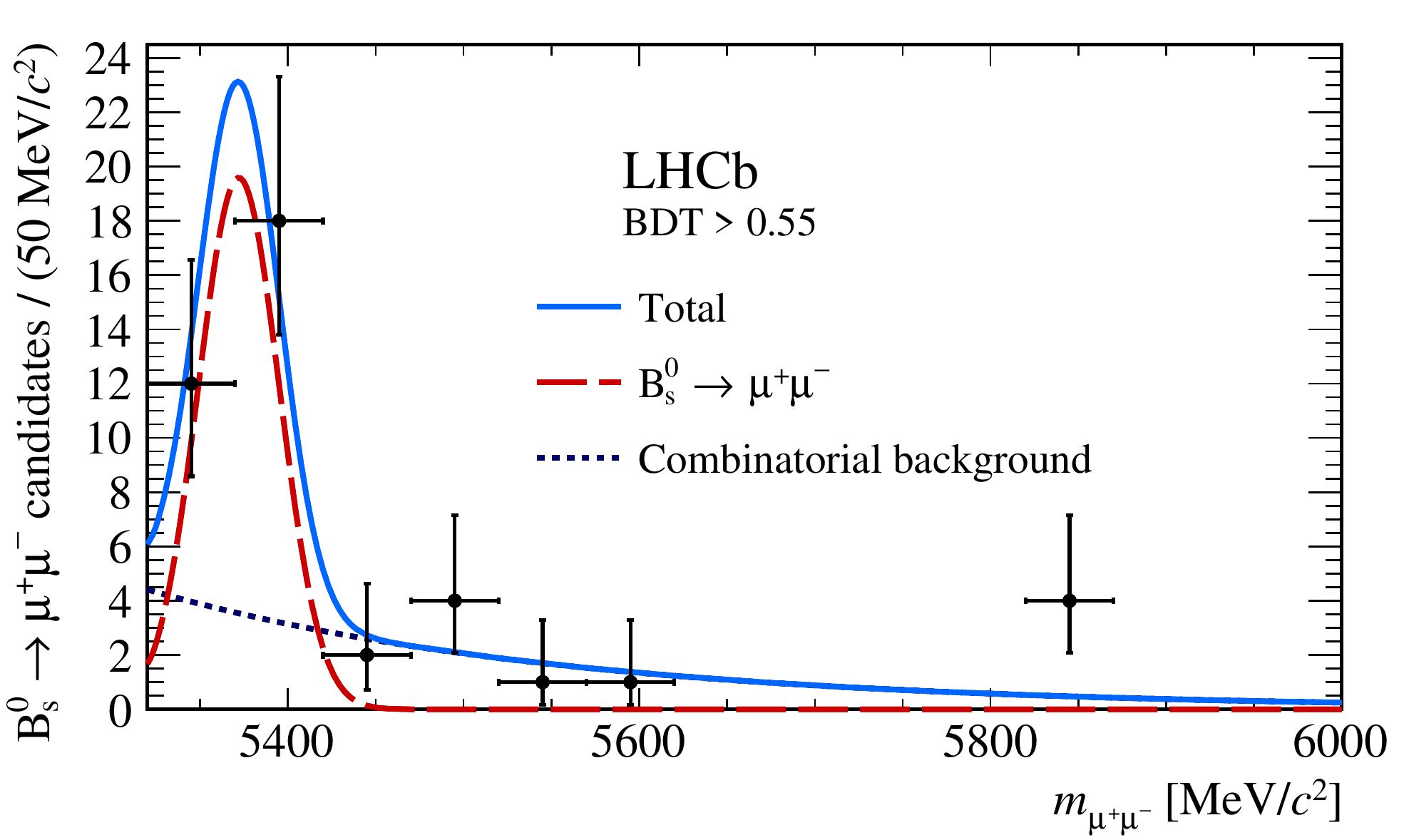}
\includegraphics*[width=.9\columnwidth]{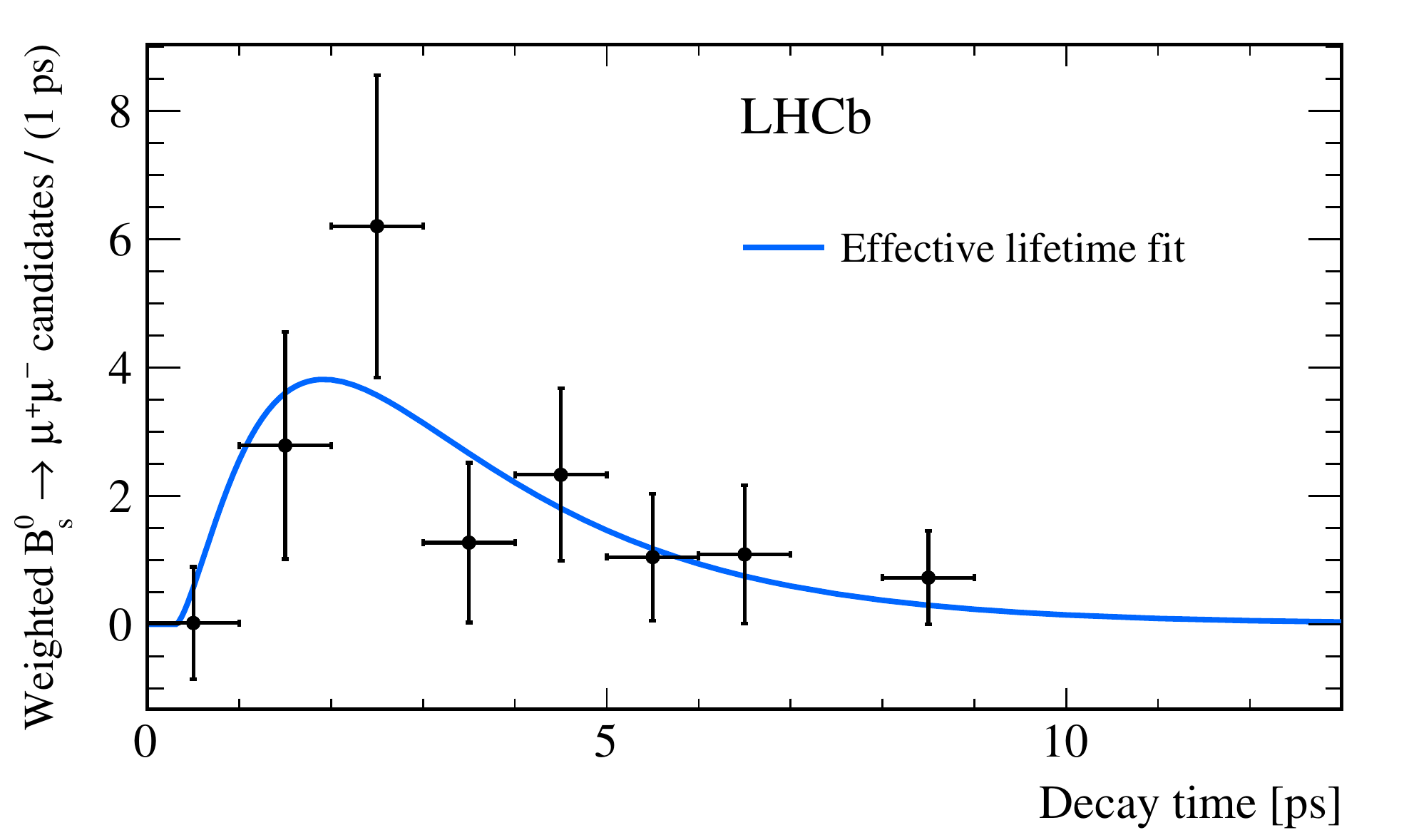}
\end{center}
\caption{\small (Top) Mass distribution of the selected \Bmm candidates (black dots) with ${\rm \BDT}>0.55$. The result of the fit is overlaid together with the \Bsmm (red dashed line) and the combinatorial background (blue dashed line) components. (Bottom) Background-subtracted \Bsmm decay-time distribution with the fit result superimposed.}\label{fig:lifetime_fit}
\end{figure}
The mass distribution of the selected \Bsmm candidates is shown in Fig.~\ref{fig:lifetime_fit} (top). Figure~\ref{fig:lifetime_fit} (bottom) shows the background-subtracted \Bsmm decay-time distribution with the fit function superimposed~\cite{roofit}. The fit results in $\tau(\Bsmm)=2.04\pm 0.44\pm 0.05 \ps$, where the first uncertainty is statistical and the second systematic. This measurement is consistent with the $\ADeltaGamma=1~(-1)$ hypothesis at the $1.0~\sigma$ ($1.4~\sigma$) level. Although the current experimental uncertainty allows only a weak constraint to be set on the value of the \ADeltaGamma parameter in the physically allowed region, this result establishes the potential of the effective lifetime measurement in constraining New Physics scenarios with the datasets that LHCb is expected to collect in the coming years~\cite{LHCb-PII-EoI}.

In summary, a search for the rare decays \Bsmumu and \Bdmumu is performed in $pp$ collision data corresponding to a total integrated luminosity of 4.4\invfb. The \Bsmumu signal is seen with a significance of \Bssigmafirst standard deviations and provides the first observation of this decay from a single experiment. The time-integrated \Bsmumu branching fraction is measured to be \mbox{\Bsbrshort}, under the $\ADeltaGamma=1$ hypothesis. This is the most precise measurement of this quantity to date. In addition, the first measurement of the \Bsmumu effective lifetime, \mbox{$\tau(\Bsmm)=2.04\pm 0.44\pm 0.05 \ps$}, is presented. No evidence for a \Bdmumu signal is found, and the upper limit \mbox{$\BRof \Bdmumu < \Bdobslimitnf$} at 95\% confidence level is set. The results are in agreement with the SM predictions and tighten the existing constraints on possible New Physics contributions to these decays.

% Do not include this in analysis note and conference reports
\section*{Acknowledgements}
\noindent We express our gratitude to our colleagues in the CERN
accelerator departments for the excellent performance of the LHC. We
thank the technical and administrative staff at the LHCb
institutes. We acknowledge support from CERN and from the national
agencies: CAPES, CNPq, FAPERJ and FINEP (Brazil); MOST and NSFC (China);
CNRS/IN2P3 (France); BMBF, DFG and MPG (Germany); INFN (Italy); 
NWO (The Netherlands); MNiSW and NCN (Poland); MEN/IFA (Romania); 
MinES and FASO (Russia); MinECo (Spain); SNSF and SER (Switzerland); 
NASU (Ukraine); STFC (United Kingdom); NSF (USA).
We acknowledge the computing resources that are provided by CERN, IN2P3 (France), KIT and DESY (Germany), INFN (Italy), SURF (The Netherlands), PIC (Spain), GridPP (United Kingdom), RRCKI and Yandex LLC (Russia), CSCS (Switzerland), IFIN-HH (Romania), CBPF (Brazil), PL-GRID (Poland) and OSC (USA). We are indebted to the communities behind the multiple open 
source software packages on which we depend.
Individual groups or members have received support from AvH Foundation (Germany),
EPLANET, Marie Sk\l{}odowska-Curie Actions and ERC (European Union), 
Conseil G\'{e}n\'{e}ral de Haute-Savoie, Labex ENIGMASS and OCEVU, 
R\'{e}gion Auvergne (France), RFBR and Yandex LLC (Russia), GVA, XuntaGal and GENCAT (Spain), Herchel Smith Fund, The Royal Society, Royal Commission for the Exhibition of 1851 and the Leverhulme Trust (United Kingdom).

% This should be taken out in the final paper

\newpage 

\addcontentsline{toc}{section}{References}
\setboolean{inbibliography}{true}
\bibliographystyle{LHCb}
%\bibliography{main,LHCb-PAPER,LHCb-CONF,LHCb-DP,LHCb-TDR}
\bibliography{main,LHCb-PAPER,LHCb-DP,LHCb-TDR}

\newpage
\centerline{\large\bf LHCb collaboration}
\begin{flushleft}
\small
R.~Aaij$^{40}$,
B.~Adeva$^{39}$,
M.~Adinolfi$^{48}$,
Z.~Ajaltouni$^{5}$,
S.~Akar$^{59}$,
J.~Albrecht$^{10}$,
F.~Alessio$^{40}$,
M.~Alexander$^{53}$,
S.~Ali$^{43}$,
G.~Alkhazov$^{31}$
% ...

{\footnotesize \it
% ...
$ ^{5}$Clermont Universit{\'e}, Universit{\'e} Blaise Pascal, CNRS/IN2P3, LPC, Clermont-Ferrand, France\\
% ...
$ ^{10}$Fakult{\"a}t Physik, Technische Universit{\"a}t Dortmund, Dortmund, Germany\\
% ...
$ ^{31}$Petersburg Nuclear Physics Institute (PNPI), Gatchina, Russia\\
% ...
$ ^{39}$Universidad de Santiago de Compostela, Santiago de Compostela, Spain\\
$ ^{40}$European Organization for Nuclear Research (CERN), Geneva, Switzerland\\
% ...
$ ^{43}$Nikhef National Institute for Subatomic Physics, Amsterdam, The Netherlands\\
% ...
$ ^{48}$H.H. Wills Physics Laboratory, University of Bristol, Bristol, United Kingdom\\
% ...
$ ^{53}$School of Physics and Astronomy, University of Glasgow, Glasgow, United Kingdom\\
% ...
$ ^{59}$University of Cincinnati, Cincinnati, OH, United States\\
% ...
}
\end{flushleft}

\clearpage
\clearpage

\end{document}